# Nuclear Criticality as a Contributor to Gamma Ray Burst Events

Robert Hayes
122 W. Permian Dr.
Hobbs, NM 88240
575-631-6264, nuclear@mit.edu

**Abstract** – Most gamma ray bursts are able to be explained using supernovae related phenomenon. Some measured results still lack compelling explanations and a contributory cause from nuclear criticality is proposed. This is shown to have general properties consistent with various known gamma ray burst properties. The galactic origin of fast rise exponential decay gamma ray bursts is considered a strong candidate for these types of events.

## Background

The Oklo uranium mine in South Africa is a place where the natural abundance of U235 was found to be consistently less than the 0.72% found everywhere else on the planet (with the remainder of the uranium being almost entirely U238). It was not until the excess isotopic distribution from the stable fission product decay nuclides was found that it was realized this was an ancient natural nuclear reactor (Cowan 1976). Because the half life of U235 is almost an order of magnitude smaller than that of U238, the ratio of U235 to U238 increases in reverse time. By going back far enough in time, the abundance of the two nuclides would become comparable. With sufficiently enriched uranium in the soil, addition of moisture from rain would then provide the necessary moderation for the earth to sustain a natural nuclear chain reaction until the heat would drive away the water and return the system to a subcritical state. This process is believed to have occurred over a billion years ago in central Africa (Fujii et al. 2000).

If the Oklo reactor occurred naturally as a result of uranium created in supernovae (SN) events, it is not incredible to think that it could occur prior to accumulation into planetary form from other SN events. In other words, just after the SN creation of the uranium and transuranic isotopes, most of the already radioactive elements created will quickly decay into stable atoms leaving these fissile and fissionable isotopes close to their original abundance. Even if a large fraction of the Pu239 (which is fissile with a moderately long half life of just over 24 ka) were to decay away, there would still be almost the initial amount of U233 (which is fissile with a moderately long half life of just under 0.2 ma) with only a negligible fraction of the U235 having decayed away. If a gravitational accumulation of the SN excreta were to occur within some few years, a longer list of potential fissile isotopes could be present in the material including Pu241,

Am243, Cm243, Cm245, Cm247, etc while the majority of the radioactive elements would still have decayed into stable isotopes. This presents a dynamic reactivity dependency in terms of the fissile nuclides generated which would be strongly dependent on the initial isotopic distribution source term.

The model presented here has the potential to explain some of the highly varied nature of a portion of measured gamma ray bursts (GRBs). These include the random nature of the events in terms of their parametric values and ranges. Specifically many measured values of energy and time distributions, pulsing, and afterglows appear to fit those attainable with the proposed criticality model.

**Introduction**

*Fundamentals of criticality*

A critical system is one in which the number of neutrons generated in one generation are just equal to the number generated in a successive generation. When a fissionable assembly is held right at the critical state, the neutron economy is largely controlled by the delayed neutrons (these come from a handful of fission products which decay by neutron emission) and so this is called a delayed critical system. These neutrons are part of the overall balance but having very long half-lives, they average in to the neutron production rate for successive generations giving the reactor a long period allowing small changes in reactivity to result in only small power changes.

The reactivity worth of the delayed neutrons at critical is referred to as 1\$ and depends primarily on the fuel mix but also to a small extent on the temperature of the system. These delayed neutrons are controlling when a reactor is at critical because they have lifetimes measured in minutes so the typical 10e-8 s lifetime of fission neutrons does not drive the period of the reactor. This delayed critical region is where nuclear power reactors are run only when increasing power output but then brought back down to the critical level for steady state operation at the higher power level. When increasing reactivity above critical to increase power, the reactivity insertion is always limited to a small fraction of 1\$ (to insure reactor operation only occurs in the range of subcritical to delayed critical and never approaching prompt critical). If the reactivity increase in a critical system were ever greater than 1\$, a doubling period $\sim 10^{-8}$ sec would occur and some nuclear yield would be expected as found in fission triggers of thermonuclear weapons. A prompt critical system can be difficult to attain mechanically because a system has to go through both the critical state and then the delayed critical range before it can reach the prompt critical range. With terrestrial systems, feedback mechanisms in commercial reactors stop or disassemble the system in the delayed critical range prior to ever reaching a prompt critical state. Typical power reactor operation occurs when a reactor is held right at the critical state. When electricity demands are constant, power levels are constant but when electricity demands decrease from a daily peak, the reactor will be brought subcritical to the lower desired power level and then brought back to the critical level to once again produce steady state power at the lower level. Conversely, in the delayed critical state, reactor power output will steadily increase. A system left in a delayed critical state will not attain prompt criticality but will rather continue to increase in power until reactivity is reduced to at least the critical level through various intentional means, thermal related shutdown or some disassembly effects driving the reactor subcritical.

In the prompt critical range, the rate of reactivity increase will exponentially determine the system disassembly feedback rate as classical thermal expansion rates tend to scale with reactivity increase below prompt critical but are negligible above prompt critical. When prompt criticality is first attained, total reactivity increase rates beyond this determine the resulting nuclear yield. This is what allows a nuclear yield to occur although obtaining a high efficiency yield is still very difficult as described by Kessler et al (2008).

Power reactor control itself is carried out by first bringing a reactor into a zero power critical state. Power is then increased slowly by raising the reactivity to a delayed state while being very careful never to approach prompt criticality. In this regime of a supercritical state, linear power increase is carried out to first order (slow increases are achieved when total reactivity is just slightly over the critical state). Likewise, reducing the power to a fixed level is done by bringing the reactor to a subcritical state until the desired lower power level is reached and then increasing reactivity again to the critical level ending in a steady state configuration (at the desired power level). Reactivity control changes are carried out by moving neutron poisons (non-fissile isotopes with very large neutron absorption cross sections) in and out of the reactor in the form of control rods.

Other parameters which effect reactivity are specific fissile isotope distributions in terms of density and moderator content with their distribution and reflector material along with the presence of any neutron poisons. All of these parameters can be modified to either increase or decrease reactivity. The most important parameter to maintain criticality is always the total fissile mass present, even if this is at a low density. In an astrophysical environment, the minimum density could even be limited by the neutron lifetime in loosely coupled systems.

In general, a neutron source is not needed to initiate a critical event if control is not desired. Most fissile nuclides already have some small spontaneous fission branching ratio for decay (KAPL 2010) and so a critical assembly left to itself long enough will initiate the chain reaction at or above critical reactivity.

*Fast and thermal fission of different isotopes (initial and time profiles)*

The initial gamma distribution from fission events depends on a number of variables and so is not expected to be the same for all fission events. One of these variables is the actual fissile radionuclide undergoing fission (Valentine 2000). In principle, the fissile abundances would depend on the age of the SN excreta and the type of SN(s) originating the material. Sneden et al describe measured hydrogen through uranium and thorium abundances for CS 22892-052 which are similar to those found in our solar system. Without any form of fractionation, this distribution is not expected to generate a critical system as a homogenous mix but it could with an adequate heterogenous mix. The standard r-process of successive neutron captures does allow for initial differential element production with the heavier elements being preferentially produced in regions of higher neutron flux (Otsuki et al. 2003).

Another parameter effecting a fission gamma distribution is the average lethargy of the neutrons causing fission (i.e., whether it is a fast neutron spectrum or a thermal neutron spectrum, Lamarsh 1983). Not only does the initial gamma distribution depend on the fissile isotope but the time profile of the delayed gamma spectrum also changes with isotope (El-Wahab 1982). The changes are not dramatic resulting in a reasonably predicable profile for a single pulse event sufficient for detection of special nuclear materials (Gozani 2009, Norman et al. 2004). Initiation of a critical configuration in an accretion disk could have a variety of these combinations present including variable assembly rates.

The key element in determining whether a system will be expected to have a thermal neutron spectrum or a fast spectrum generally only depends on moderator content (typically hydrogen). This is because the minimum critical mass for thermal (moderated) systems tends to be more than an order of magnitude lower than that for a fast system. This in turn is because the fission cross section is often many orders of magnitude larger for slow neutrons than for fast neutrons. If a system has adequate moderation present intermixed with the fissile materials, far less fissile material is needed to be accumulated for a critical configuration.

*Nuclear criticality coupling (geometry, buckling or material distribution)*

Any critical system can be cut into multiple pieces resulting in multiple subcritical systems. In the simple case of cutting a critical system in half, as the two systems are brought back together, reactivity increases as a result of the interaction between the two systems. In general, any critical system could be made supercritical by contracting its fissile material (as is done with bombs) although for thermal systems, this would also require maintaining adequate hydrogen (moderator) content. The coupling can readily be made more complicated by changing the geometry, density gradients, interstitial material between reactive regions and reflector distribution. Depending on the change, any of these perturbations could increase or decrease system reactivity accordingly.

Weakly coupled thermal systems (typically spatially large systems) can also undergo xenon oscillations (Eliasi et al 2011) because Xe135 is a fission product poison (FPP). These oscillations occur when regions of higher fission density (initially at the center of a reactor core) create proportionately more fission product poisons than neighboring low fission density regions. This sets up an oscillation whereby the less poisoned regions experience less FPP production and so undergo a relative increase in fission rate such that when the previous FPPs are burned up, the process repeats with FPPs always being both produced and consumed in proportion to the thermal neutron flux. In a thermal reactor, neutron flux is proportional to both fission rate and power density which drives these nominal peak flux location dynamics. In fast systems, FPP is not a large effect because the neutron absorption cross sections of the FPP's are only large for thermal neutrons.

*Criticality approximation*

The overall neutron balance for a delayed critical system occurs when the multiplication factor $k$ (the ratio of fissions in any given generation to its subsequent generation) is unity. This can be estimated for an infinite thermal reactor by $k_\infty = \eta\, f\, p\, \varepsilon$ where $\eta$ is the number of free neutrons generated per neutron absorbed by the fuel, $f$ is the fraction of thermal neutrons absorbed in the fuel, $p$ is the probability that a neutron will reach thermal energy, and $\varepsilon$ is the ratio of the fast to thermal neutron population (Culp 1991).

It can be shown that the thermal neutron non leakage probability can be approximated by $P = 1/(1+\lambda^2 B^2)$ and that the fast nonleakage probability can be approximated by $F_{nl} = \exp(-B^2 \tau)$. Here, $\tau$ is the neutron age (in units of area), $\lambda$ is the thermal diffusion length and $B$ is the geometric buckling. These values for standard water are $\tau \cong 27$ cm$^2$, $\lambda \cong 3$ cm and for compact reactor shapes of characteristic radius $R$, $B \cong \pi/R$. When reactor density is changed, the following approximations are used $\tau \sim \tau_0\, (\rho_0/\rho)$, $\lambda \sim \lambda_0\, (\rho_0/\rho)\, (T/T_0)^{1/2}$ (Lamarsh 2002) relative to standard conditions. With this, a critical system can be approximated to first order by $k \sim \eta f p\, \varepsilon\, F_{nl}(R, \rho)\, P(R,\rho,T)$ as given in equation 1

$$k \approx \eta\, f\, p\, \varepsilon\, \exp\!\left(-\tau_0 (\rho/\rho_0)(\pi/R)^2\right) / \left(1 + (\pi/R)^2 (\rho_0/\rho)^2\, T/T_0\right) \qquad \text{Eq. 1}$$

Other approximations can be used for alternate moderating materials and configurations but with the current sophistication available in Monte Carlo codes (Sood et al. 2003), almost all modern reactivity calculations are done on the computer.

*Accidental criticalities*

Inadvertent criticalities have been well characterized and studied over the years under a wide variety of industrial settings (Knief 1985). One common occurrence is the pulsing of the system repeatedly going critical over time. This is sometimes caused when the accidental criticality heats the system to the point where the lowered density drives the system subcritical due to increased neutron leakage as both $F_{nl}$ and $P$ decrease with density reduction. In such cases, after cooling and reassembly, the system goes critical again at the higher density. The trend so far has always been that subsequent pulses are smaller than the first. This is attributed to the combined effects of addition of fission products to the system (which are partially composed of neutron poisons) in addition to ejected fissile content (in some cases). Both properties appear common not only to inadvertent criticalities but also GRBs in terms of energy output dynamics. This tendency for some gamma-ray bursts to pulse in diminishing magnitudes (Barbiellini and Longo 2003) might be the result of a critical mass occurring in an accretion disk (or other gravitational concentrator) driving itself subcritical by the thermal expansion from the criticality event and then reassembling at a later time in a repetitive manner but with additional fission product poisons added from previous pulses.

It is this very criticality pulsing phenomenon that drives some nuclear facilities to require criticality alarm systems to alert workers to leave the area due to the possibility of later criticality events taking place after the initial detected pulse (ANS 2003).

Fission criticality has been considered as a possible contributor to r-process SN (Qian 2003) but not as a contributor to GRB events. Both processes would require a sufficient quantity of actinides present.

*Actinide genesis and evolution*

Current model requirements to attain the isotopic distributions for actinides found in our solar system have not been made to fit observations better than a factor of 2 for both the actinides (Wasserburg et al. 2006) and non-actinides (Langanke 1999). Progress in explaining observations is continually ongoing (Nomoto et al. 2006) with fission readily recognized as being an important cutoff mechanism in r-process actinide growth process (Martinez-Penedo et al. 2007).

Standard models of chemical fractionation from IS media and SN excreta include the ELS model which has a proto-galaxy undergoing a rapid collapse followed by the remaining gas forming a metal rich cold disk (Gilmore et al. 1998). Here thermal dissipation of heat during a gravitational collapse is coupled with differential thermal conductivities of chemical species and angular momentum combine to provide elemental fractionation. In this way, the most metal poor stars end up with the lowest angular momentum orbits in a galaxy. Chemical fractionation of lighter elements and in particular, smaller organic molecules seem to be quantitatively predictive based on current models (Dishoeck and Blake 1998) although no detailed chemical dynamics for heavy metals is currently able to produce highly compelling results.

Other models considering actinide genesis include neutron star mergers (Goriely et al. 2005) and neutron star, black hole mergers (Domainco and Ruffert 2005). This process alone would generate substantial gamma ray spectral lines from an abundance of transuranics generated but not necessarily the criticality event prompt gamma distribution directly followed by a delayed fission product decay gamma spectra. Mixing of this material into a subsequent critical mass with (or without) hydrogen in an accretion disk would generate this pattern of gamma rays as described below.

**Results and Discussion**

*GRB Energy Spectra*

If the fissile material were concentrated in an accretion disk, a rapid density increase above the critical level can be postulated. This is because neutrons from successive generations have to spread out from the initial critical event to all other regions capable of sustaining criticality and if this region is small, strong coupling is expected. Any weak coupling between regions can cause temporal variations which can be as symmetric as the reactor or highly asymmetric if a critical cloud were extant. If the regions are separated by spaces not capable of supporting a fission chain reaction then only those neutrons streaming back and forth through these dead spaces would be able to communicate successive neutron generation information. Specifically if two separate regions are individually subcritical (but critical when unified) and continuously brought near each other, a complex time evolution becomes possible (particularly when it is not just two symmetric systems being assembled). It is possible then that heating from fission could expand the system to subcriticality followed by successive gravitational coalescence to regain the critical state in a repetitive fashion, particularly if the criticality were occurring in an accretion disc.

The characteristic energy distribution for GRBs has been described as that given by Equation 2, where $A$ is an arbitrary magnitude parameter and $\beta$ tends to vary between around 1 or 2. More specifically, the $\beta$ exponent in Equation 2 from BATSE (Waxman 2003, Hurley 1989) has been estimated as $\beta \sim {}^-2$ when the energy $E$ is in the range of 0.02-2 MeV with $\beta$ values at higher energies being $\beta \sim {}^-1$ (although there are a number of reasonable ranges around these parameters proposed elsewhere, Mészáros 2002).

$$N(E) \propto A e^{-\beta E} \qquad (2)$$

The implications from Equation 2 become significant when compared to the characteristic energy dependence of prompt and delayed gamma emissions from fission events[†] (Hayes 2009). The functional form for generic prompt fission gamma rays is given in Equation 3 (Shultis and Faw 2000). The initial distribution for the prompt gamma emissions seems to follow reasonably well with those seen from BATSE GRB observations described above making for a compelling plausibility argument of the fission contribution to different GRB events.

$$N_{prompt}(E) = \begin{cases} 6.6 & 0.1 < E < 0.6 \text{ MeV} \\ 20.2 e^{-1.78E} & 0.6 < E < 1.5 \text{ MeV} \\ 7.2 e^{-1.09E} & 1.5 < E < 10.5 \text{ MeV} \end{cases} \qquad (3)$$

The functional form for delayed fission energy distribution is given by Equation 4 (Shultis and Faw 2000). These are the fission product decay gamma emissions given off after the critical chain reaction has been stopped. The parameter $m(t)$ in Equation 4 has a magnitude dependent on the total number of fissions and a known time dependence. The expected GRB afterglow energy distribution is approximated by $\beta \approx {}^-1$ (Paradijs et al. 2000) which again is in very nice agreement with the spectral shape of the delayed gamma photon distribution of a fission criticality event given in Equation (4).

$$N_{delayed}(E,t) \approx m(t) e^{-1.1E} \qquad (4)$$

So both the prompt and delayed GRB energy distributions are in rather remarkable agreement with the spectral shape expected from fission events. This provides a very compelling argument to say that IS criticality events are likely contributing to measured GRB phenomena.

*GRB time evolution*
Less agreement is found in the time profile $m(t)$ of the delayed fission gamma distribution predicted by Ghisellini (2003) when compared with average gamma ray burst after glows. This was approximated as having a roughly $m_{GRB}(t) \sim t^{-1/4}$ dependency for the first few seconds although Paradjis et al (2000) put this dependence as $m_{GRB}(t) \sim t^{-1.15}$ for numerous afterglow events. The delayed gamma distribution from fission has been shown to sometimes have one or more knees with the initial functional dependence generalized by Equation 5 (Tobias 1980). Comparison with the profile given by Paradjis of the time exponent is in reasonable agreement with the fission criticality contributions to GRBs model.

---

[†] Note that prompt gamma photons are those emitted during the fission process, delayed emissions are those emitted from the decay of the fission products themselves.

$$m(t) \approx 1.26 t^{-1.2} \tag{5}$$

The actual dependence of the knees and initial slope depends on both the isotopes contributing to criticality and the burn history. As such, more accurate approximations for long burn histories are generally given by case specific series summations of decaying exponentials. The presence of any knee in a decay series can be accounted for by recognizing that some isotope decay chains can have radioactive progeny with higher specific gamma activity than a parent nuclide. Other models have been proposed which tend to adhere to similar power law decay approximations (Zhang 2007). Even with all this, there are additional bumps and wiggles (Maienschein et al. 1958) in these fission decay responses due to co-irradiated materials which can depend on interstitial materials (e.g., moderators and poisons), shielding and reflectors. In this sense, agreement with the exponent in equation 5 would not be expected to be as strong as those in equations 2-4.

Of particular note here is that these spectra are consistent with the fast rise exponential decay (FRED) events described by Tello et al. (2012). These events are argued by Tello to be GRB sources localized to our galaxy which do not appear to be excluded in any way from the criticality model proposed here and so are hypothesized to include nuclear criticality source terms.

*Beaming mechanisms*

The seminal experiments by Wu et al (1957) showed that a sufficiently strong magnetic field can polarize Co60 nuclei which would then emit electrons through radioactive decay. Electron emission like this by radioactive decay has been known to have angular correlation to gamma emission for some time (Stevenson and Deutsch 1951, Reddinguis et al. 1968, Reddinguis and Postma 1969, Avignone et al. 1965 etc). With the majority of fission products being beta decay gamma emitters, preferential emission angle biasing of the beta and gamma emissions would be expected. Sufficient coupling of a large external magnetic field and a fission product nucleus could provide a mechanism for limited beaming gamma rays from a nuclear criticality event in space if the event took place in the immediate presence of a magnetar (Hurley 2011). The biased electron emission beaming due to the magnetic field would initiate the coupling but it would also be expected to complicate the expected gamma ray spectrum measured due to the differing coupling from each of the individual fission product gamma emitters (Equations 2 through 5 are measurements for all isotopes not taken in a magnetic field).

Nuclear polarizations near 25% in terrestrial systems based on lowering temperatures in an external magnetic field ($10^5$ G) are considered attainable as given in Equation 6 taken from Roberts and Dabbs (1961). The terms used in Equation 6 are; $P$ = nuclear polarization, $I$ = nuclear spin, $\mu_n$ = nuclear moment, $H$ = magnetic field strength, $T$ = temperature in Kelvin with all terms other than $H$ and $T$ being taken to be on order of unity here. The equation neglects hyperfine interactions but clearly places the potential to have almost unity polarization for most nuclides in the field ranges of magnetars at $10^{14}$ G (Hurley 2011) where temperatures can be on the order of $10^6$ K. This gives a substantial range of temperatures and field strengths to still realize large polarizations for most radionuclides. Prompt fission gamma correlation has even been measured to be dependent on fission fragment directions (Fraser and Milton 1966) with the measured anisotropy in fission gamma rays being largely independent of decay gamma energy (Kandil and El-Mekkawi 1974).

$$P = 1.22 \times 10^{-8} (I+1) \mu_n H / (I T) \qquad (6)$$

The resultant gamma anisotropy from a polarized radionuclide can therefore have many parametric dependencies for any gamma beaming including decay energy, radionuclide spin state, external magnetic field strength and system temperature. Although this does nothing more than provide a possible beaming bias, general considerations still place the source of these events within our galaxy.

To the extent that the magnetar spin axes are highly correlated with the galactic axis, the gamma anisotropy relative to the galactic plane could be biased to preferentially orient fission product decay gamma emission along the galactic axis. Such an emission mechanism would give a more homogenous distribution appearance from purely galactic sources. The extent to which magnetars in the galactic halo have spins preferentially aligned with the galactic axis is indeterminate but qualitatively viable.

*Contraindications for cosmogenic fission GRB*

Most long-duration GRB have been demonstrated to be located in the high redshift universe. At such distances, the GRB energies are attributed with energies from 1e50 to 1e53 ergs. There is a host of compelling theory coupled with both direct and indirect evidence establishing the source of GRB as the death of massive stars (Woolsey and Bloom, 2006). The distance range of these objects extends out beyond $z \sim 0.01$. An example derivation for ascribing this distance scale is provided below.

When electromagnetic radiation interacts with matter, the dominant mechanism is scatter off of electrons. The oscillating electric field in the photon jiggles the electron which then generates a scatter photon with the electron being kicked away in recoil. For instance, a gamma-ray burst of typical duration $t \sim 1$ min must originate from a region of characteristic size $L < t*c \sim 10^{12}$ cm, since otherwise its duration would be increased due to light crossing time delay (this is actually a conservative upper limit on the emission size, since what truly matters is the variability timescale, which is typically $\sim 1$ second or less). In addition to being smaller than the light crossing size, the emission region must be transparent to gamma-rays. The requirement that the transmission region must be transparent to gamma-ray photons can be translated into a constraint on the optical depth $\tau_\varepsilon = n_e * \sigma_T * L < 1$ to Thomson scattering, where $\sigma_T = 6.995*10^{-25}$ cm$^2$ is the Thomson scattering opacity; $n_e \sim N_e / L^3$ is the average number density of electrons and $N_e$ is the total number of electrons. Requiring both $\tau_\varepsilon < 1$ and $L < 1e12$ cm implies that $N_e < 10^{48}$. Assuming fully ionized gas, the number of electrons $N_e$ is similar to the number of nucleons $N_n$.

If every fission reaction optimistically released $E_{fiss} \sim 1$ MeV per nucleon, then this places an upper limit of $E_{tot} < E_{fiss} * N_e \sim 10^{42}$ erg. Given that GRBs have typical fluence of $\Phi \sim 10^{-6}$ erg/cm$^2$, such an event would have to originate from a typical distance D $\sim \sqrt{(E_{tot} / \Phi * 4 * \pi)} \sim 100$ kpc or less, i.e. in our Galaxy or its halo.

*Predicted evidence from the fission model*

There are yet a large number of unknowns which can vary in the model proposed. That the criticality model has sufficient parametric latitude to cover much temporal and energy variation seen in gamma-ray bursts is not much more than a compelling demonstration of consistency. Also, there are some very unique spectral signatures which could be attributed to a criticality event as measured in terrestrial fission events (Hoffman and Hoffman 1974, Rudstam et al. 1990). One of these gamma signatures is the 2.2 MeV gamma-ray emitted by thermal absorption of a neutron by hydrogen (applicable to fission events sustained by hydrogen thermalized neutrons). As this is not accompanied by a beta emission, beaming would depend on polarization of the neutron and proton prior to absorption (Müller et al 2000).

If a fast system were the source of a particular gamma ray burst, there are unique beta related decay events which should be present in the initial burst and afterglow depending on the isotopes present in the neutron field. This could be a long list depending on the particular mix but would offer confirmatory evidence if found. Perhaps one of the more expected of these is the fast neutron scatter event on O16 generating N16 and a proton, n(O16,N16)p because the N16 decays with a 7 sec half life through beta emission and a very hard 6.1 MeV gamma and (with a lower branching ratio) 7.1 MeV gamma or the n(Fe,n)p gamma from iron if sufficient iron were present in the system. Some measurements have already verified radioactive isotopes in supernovae excreta through gamma spectroscopy (Clayton 1992, Leising 2006) so additional predictions of specific fission signatures from some GRB and their afterglows are expected from this model.

When sub microsecond resolution of the initial GRB become available, these should also have GRB events consistent with that measured from terrestrial fission. Some differences between fast systems and thermal systems are scaled on the neutron lifetime which is on the order of a μsec for thermal systems (Maier-Leibnitz and Springer 1966) and around two orders of magnitude smaller for fast reactors (Koch and Paxton 1959).

*Galactic GRB*

Current measurement results largely exclude nuclear criticality as deriving from any cosmological origin. There is a small subset of galactic events known as fast rise exponential decay (FRED) events described by Tello et al (2012) which could be attributed to nuclear criticality events as proposed.

Another source of galactic GRB are the soft gamma repeaters. These sources have been described as magnetars undergoing plate techtonic like magnetic stresses leading to Alfvèn wave injection to the magnetosphere ultimately resulting in a measurable flare (Kouveliotou et al 1998). Cooper and Kaplan (2010) also describe magnetar behavior supporting an SGR source explanation. In the latter work, magnetic field losses are postulated to reduce crustal integrity causing electron captures. These events do not have the same spectral energy distribution given in Equations 2 and 3 and so are not consistent with the fission model.

**Conclusions**

Even though the details of r-process production of the fissionable isotopes are not understood in overwhelming detail (Arnould et al. 2007), all that is required for the current model is some form of chemical or genesis fractionation of the actinides (preferably in the vicinity of a magnetar to promote beaming). The most likely candidate identified for these events are the FRED GRB's described by Tello et al. (2012) having galactic origin.

Given these assumptions, it has been shown that interstellar nuclear criticality events are a credible contributor to measured gamma ray bursts. The supporting evidence includes initial burst gamma energy distribution up to around 10 MeV, the random after pulsing and the time series of many afterglow gamma distributions. Additional predicted evidence for thermal fission events would be a 2.2 MeV gamma ray from hydrogen absorption of a thermal neutron n+p->d and for a fast fission event, the various telltale elastic and inelastic scatter gamma rays produced from neutron interactions.

*Acknowledgements* Special thanks are extended to Dr Peter Fisher of the MIT physics department's laboratory for nuclear science for useful discussion and encouragement in consideration of this work.